\documentclass[prb,twocolumn,showpacs,preprintnumbers,amsmath,amssymb]{revtex4}

\usepackage{amsfonts}
\usepackage{latexsym}
\usepackage{graphicx}
\usepackage{dsfont}
\usepackage{epsfig}

\newcommand{\be}{\begin{equation}}
\newcommand{\ee}{\end{equation}}
\newcommand{\cuo}{\ensuremath{\textrm{CuO}_2}}
\newcommand{\cud}{\ensuremath{\textrm{Cu~3d}_{x^2-y^2}}}

\newcommand{\tsp}{\ensuremath{t_{\mathrm{sp}}}}
\newcommand{\tpd}{\ensuremath{t_{\mathrm{pd}}}}
\newcommand{\tpp}{\ensuremath{t_{\mathrm{pp}}}}
\newcommand{\tss}{\ensuremath{t_{\mathrm{ss}}}}
\newcommand{\epp}{\ensuremath{\epsilon_{\mathrm{p}}}}
\newcommand{\epd}{\ensuremath{\epsilon_{\mathrm{d}}}}
\newcommand{\eps}{\ensuremath{\epsilon_{\mathrm{s}}}}
\newcommand{\ep}{\ensuremath{\varepsilon_{\mathrm{p}}}}
\newcommand{\ed}{\ensuremath{\varepsilon_{\mathrm{d}}}}
\newcommand{\es}{\ensuremath{\varepsilon_{\mathrm{s}}}}

\begin{document}

\title{LCAO model for 3D Fermi surface of\\
high-$T_c$ cuprate ${\rm Tl_{2}Ba_{2}CuO_{6+\delta}}$}

\author{M.~Stoev}
\email[E-mail: ]{martin.stoev@gmail.com}
\author{T.M.~Mishonov}
\email[E-mail: ]{mishonov@phys.uni-sofia.bg}

\affiliation{Department of Theoretical Physics, Faculty of Physics,\\
University of Sofia St.~Kliment Ohridski,\\
5 J. Bourchier Boulevard, BG-1164 Sofia, Bulgaria}

\date{\today}

\begin{abstract}
A simple analytical formula for three-dimensional Fermi surface
(3D FS) of ${\rm Tl_{2}Ba_{2}CuO_{6+\delta}}$ is derived in the
framework of LCAO approximation spanned over Cu~4s,
Cu~3d$_{x^2-y^2}$, O~2p$_x$ and O~2p$_y$ states. This analytical
result can be used for fitting of experimental data for 3D FS such
as polar angle magnetoresistance oscillation. The model takes into
account effective copper-copper hopping amplitude \tss\ between
Cu~4s orbitals from neighbouring \cuo\ layers. The acceptable
correspondence with the experimental data gives a hint that the
\tss\ amplitude dominates in formation of coherent 3D FS, and
other oxygen-oxygen and copper-oxygen amplitudes are rather
negligible. For absolute determination of the hopping parameters a
simple electronic experiment with a field effect transistor type
microstructure is suggested. The thin superconductor layer is the
source-drain channel of the layered structure where an AC current
is applied.
\end{abstract}



\maketitle
\section{Introduction and notations}
%
For a long time magnetic oscillations are a standard method for
determination of  Fermi surface. For a comprehensive introduction
see for example the monograph by Shoenberg \cite{Shoenberg}. A
recent observation of three-dimensional Fermi surface (3D FS) in
${\rm Tl_{2}Ba_{2}CuO_{6+\delta}}$ (Tl:2201) \cite{Hussey}
unambiguously has shown that charge carriers in this material are
ordinary Fermi particles. This observation has an important
significance for the physics of high-$T_c$ cuprates in general.
There is almost a consensus that pairing mechanism is common for
all cuprates wherever it is hidden. That is why the observation of
3D FS leads to the conclusionthat this mechanism should be able to work even for
Fermi quasiparticles in BCS scenario. An important first step in
this scenario is the analysis of the electron spectrum of a metal in a
selfconsistent approximation of independent electrons. For the
cuprates, as for many other ion materials, the method of linear combination of
atomic orbital (LCAO) gives adequate description of the band
structure. Tight binding band structure is also a relevant
starting point for theoretical analysis of many phenomena related
to quasiparticle interaction. For example
Abrikosov~\cite{Abrikosov} used tight binding $\sigma$-model to
build the spin density theory of metal-insulator transition in
cuprates. A lot of phenomena in cuprate physics especially for
overdoped cuprates can be understood in the framework of orthodox
fermiology. The purpose of the present work is to derive analytical LCAO formula
for 3D FS and the band structure of Tl:2201 which can be used for
further analysis of the experimental data; for local density
approximation (LDA) calculations of band structure of this
material see, for example, reference \cite{Singh}. These
analytical results can be used for fitting the angle
magnetoresistance oscillation (AMRO) data and at the same time
they represent a realistic noninteracting part of the lattice
Hamiltonian for further consideration of pairing in cuprates.

Hilbert space in LCAO approximation is spanned over the relevant
Cu~3d$_{x^2-y^2}$, Cu~4s, O~2p$_x$ and O~2p$_y$ orbitals. The
generic 4-band $\sigma$-model for CuO$_2$ plane was suggested by
Labb\'e and Bok~\cite{LabbeBok}. Later on detailed {\it ab initio}
calculations of band structure of layered cuprates by Andersen
{\it et al.}~\cite{Andersen} confirmed that this generic model
adequately interpolates the LDA band structure and the influence
of $\pi$-orbitals for the conduction band is negligible. In
Tl:2201 the neighbouring \cuo\ planes are shifted in a half
period. In an elementary cell indexed by three integer numbers
${\rm {\bf R_n}={\bf a}_1 n_1 + {\bf a}_2 n_2 + {\bf a}_3 n_3}$,
where ${\bf a}_1=a_0 {\bf e}_x$, ${\bf a}_2=a_0 {\bf e}_y$ and
${\bf a}_3=b_0 {\bf e}_z+\frac{1}{2}a_0 ({\bf e}_x+{\bf e}_y)$,
the space vector of copper ions  is ${\bf R}_{\rm Cu}={\bf 0}$,
and for oxygen ions we have ${\bf R}_{\rm O_a}=\frac{1}{2}a_0 {\bf
e}_x$ and ${\bf R}_{\rm O_b}=\frac{1}{2}a_0 {\bf e}_y$;\;${\bf
e}_x$, ${\bf e}_y$ and ${\bf e}_z$ are the unit vectors. For the
introduced notations the LCAO wave function reads as
\begin{equation}
\begin{array}{ll}
  \psi_{\rm LCAO}({\bf r})=\sum_{\bf n}
  [ D_{\bf n}\psi_{_{{\rm Cu 3d}}}
  ({\bf r}-{\bf R}_{\bf n}-{\bf R}_{\rm Cu})\\
  & \\
  +S_{\bf n}\psi_{_{ {\rm Cu}4s}}({\bf r}-{\bf R}_{\bf n}-{\bf R}_{\rm Cu})
  + X_{\bf n}\psi_{_{ {\rm O_a}2p_x}}
  ({\bf r}-{\bf R}_{\bf n}-{\bf R}_ {\rm O_a})\\
  & \\
  +Y_{\bf n}\psi_{_{ {\rm O_b}2p_y}}
  ({\bf r}-{\bf R}_{\bf n}-{\bf R}_{\rm O_{b}})] ,\\
\end{array}
\end{equation}
where $\Psi_{\bf n }=(D_{\bf  n},S_{\bf n} ,X_{\bf n},Y_{\bf n})$
is the tight-binding wave function in lattice
representation~\cite{MishonovPenev}. In second quantisation
approach the complex amplitudes $D_{\bf  n},S_{\bf n} ,X_{\bf n}$
and $Y_{\bf n}$ become Fermi annihilation  operators.

The LCAO wave function can be expressed by Fermi operators in
momentum space
\begin{equation}
\Psi_{\bf n}\!=\!\left(\!\
  \begin{array}{c}
    D_{\bf n}\\
    S_{\bf n} \\
    X_{\bf n} \\
    Y_{\bf n} \\
\end{array}
\!\right)\! =\! \frac{1}{\sqrt{{\cal N}}} \sum_{\bf k}e^{\rm i{\bf
k}\cdot{\bf R_n}} \left(\!\
  \begin{array}{c}
    D_{p} \\
    S_{ p} \\
    -i e^{\rm {\bf k} \cdot {\bf {R_{\rm O_a}}}} X_{p} \\
    -i e^{\rm {\bf k} \cdot {\bf {R_{\rm O_b}}}}Y_{p}
\end{array}
\!\right),
\end{equation}
where ${\cal N}$ is the number of unit cells supposing periodic
boundary condition, ${\bf k}=(p_x/a_0, p_y/a_0, p_z/b_0)$ is the
electron quasimomentum and  ${\bf p}$ is the dimensionless
quasimomentum  in first Brillouin zone $(p_x, p_y, p_z)\in
(-\pi,\pi)$. This equation  describes the Fourier transformation
between the coordinate representation $\Psi_{\bf n}=(D_{\bf
n},S_{\bf n},X_{\bf n},Y_{\bf n})$ and the momentum
representations $\psi_p=(D_p,S_p,X_p,Y_p)$ of the tight-binding
wave function.

In order to derive the electron band Hamiltonian in momentum
representation, we have to analyze the LCAO Schr\"{o}dinger
equation $\epsilon\psi_i=\epsilon_i\psi_i-\sum_j (\pm)t_{ij}
\psi_j$ for a plane wave $\Psi_{\bf n}\propto e^{i{\bf k}\cdot{\bf
R}_{\bf n}}$, where $\psi_i$ are amplitudes multiplying atomic
orbitals, $\epsilon$ is the electron energy, $\epsilon_i$ are
single site energies, the sum is spanned to the nearest and next nearest
neighbouring atoms at i-orbital and hopping amplitudes have +sign
for bonding, and -sign for antibonding orbitals with different
signs of atomic wave functions~\cite{Feynman}. For a detailed
analysis, notations and references see \cite{MishonovPenev}. Here
we will give a brief description. Let $\epsilon({\bf p})$ be the
electron band dispersion;~\epd\ is the single-site energies for
Cu~3d level,~\tpd\ is the hopping amplitude between the O~2p
states and Cu~3d. Considering only nearest-neighbour hoppings and
omitting the common $e^{i{\bf k} \cdot {\bf R}_{\bf n}}$
multiplier for the Cu~3d$_{x^2-y^2}$ amplitude in ${\bf R}_{\bf
n}$ elementary cell we obtain
\begin{equation}
\begin{array}{lr}
  \epsilon({\bf p})D_{p}\!=\!
  \epd D_{p}\!-\!i\tpd( e^{i\frac{p_{x}}{2}}X_{p}\!-\! e^{i\frac{p_{x}}{2}}e^{-ip_{x}}X_{p}
  \!-\! e^{i\frac{p_{y}}{2}}Y_{p}+\\
  &\\
  e^{i\frac{p_{y}}{2}}e^{-ip_{y}}Y_{p})
=\epd D_{p}+2\tpd(\sin\frac{p_{x}}{2}X_{p}
 -\sin\frac{p_{y}}{2}Y_{p})=\\
 & \\ = \epd D_{p}+\tpd s_{x}X_{p}-\tpd s_{y}Y_{p}.\\
\end{array}
\end{equation}
Analogous consideration can be applied for Cu~4s amplitude: $\eps$
is the single site energy for Cu~4s state, \tsp\ is the hopping
amplitude between O~2p and Cu~4s; \tsp\ is bigger than \tpd\
because \cud\ orbitals are much more localized near to copper
nucleus. Due to the symmetry of the Cu~4s wave function, now the
relative signs between $X_p$ and $Y_p$ are equal
\begin{equation}
\begin{array}{ll}
  \epsilon({\bf p})S_{p}= \eps S_{p}+\tsp s_{x}X_{p}+\tsp s_{y}Y_{p}\\
  &\\ +i\tss(e^{i\frac{p_{x}}{2}}e^{i\frac{p_{y}}{2}}e^{i{p_{z}}}
  +e^{i\frac{p_{x}}{2}}e^{i\frac{p_{y}}{2}}e^{-i{p_{z}}}
  + e^{-i\frac{p_{x}}{2}}e^{i\frac{p_{y}}{2}}e^{i{p_{z}}}\\
  &\\  +e^{-i\frac{p_{x}}{2}}e^{i\frac{p_{y}}{2}}e^{-i{p_{z}}}
  +e^{-i\frac{p_{x}}{2}}e^{-i\frac{p_{y}}{2}}e^{i{p_{z}}}
  + e^{-i\frac{p_{x}}{2}}e^{-i\frac{p_{y}}{2}}e^{-i{p_{z}}}\\
  &\\  +e^{i\frac{p_{x}}{2}}e^{-i\frac{p_{y}}{2}}e^{i{p_{z}}}
  +e^{i\frac{p_{x}}{2}}e^{-i\frac{p_{y}}{2}}e^{-i{p_{z}}})S_{p}=\\
&\\=\eps S_{p}+\tsp s_{x}X_{p}+\tsp s_{y}Y_{p}-t_{ss}zS_{p},\\
\end{array}
\end{equation}
where $z({\bf p})=c_x c_y c_z.$ The influence of three
dimensionality (3D) is taken into account only here by the
effective Cu~4s--Cu~4s transfer integral \tss\ between copper ions
from different \cuo\ planes. Following \cite{Andersen} we have
used the standard notations:
\begin{equation}
  \begin{array}{ll}
  s_x = 2\sin(p_x/2),\!&\!s_y =2\sin(p_y/2), \\
  &\\c_x = 2\cos(p_x/2),\!&\!c_y = 2\cos(p_y/2),\\
  &\\x=\sin^2(p_{x}/2),\!&\! y=\sin^2(p_{y}/2), \\
\end{array}
\end{equation}
adding $c_z= 2\cos p_z$. In first Brillouin zone for $p_x,p_y
\in (-\pi,\pi)$ the variables $c_x, c_y\ge 0,$ however in the
whole momentum space in the interval $(0,2\pi),$ for example, we
have to redefine $c_x = 2|\cos(p_x/2)|$ and $c_y =
2|\cos(p_x/2)|.$

Analogously considering electron hopping to O~2p$_x$ orbital and
dividing by $-ie^{i\frac{p_{x}}{2}}$ we have
\begin{equation}
\begin{array}{ll}
\epsilon({\bf p})X_{p}=\epp X_{p}+\tpd s_{x}D_{p}+\tsp s_{x}S_{p}+\\
&\\
+\, \! \tpp \! (-i) \!
(e^{-i\frac{p_{x}}{2}}e^{-i\frac{p_{y}}{2}}e^{-i\frac{p_{x}}{2}}e^{i\frac{p_{y}}{2}}
 \! + \! e^{i\frac{p_{x}}{2}}e^{i\frac{p_{y}}{2}} \!- \! e^{i\frac{p_{x}}{2}}e^{-i\frac{p_{y}}{2}})Y_{p}\\
&\\
=\epp X_{p}+\tpd s_{x}D_{p}+\tsp s_{x}S_{p}-\tpp s_{x}s_{y}Y_{p},\\
\end{array}
\end{equation}
where \epp\ is O~2p single-site energy, \tpp\ is the hopping
amplitude  between adjacent O~2p$_x$ and  O~2p$_y$ orbitals. Due
to the crystal symmetry  the equation for $Y_p$ can be obtained by
exchange between x and y, and X and Y; only the relative sign between
O~2p and Cu~3d orbitals has to be changed.

Finally the LCAO Schr\"{o}dinger equation in momentum
representation reads as
\begin{equation}
(H_p-\epsilon({\bf p})\mathds{1})\psi_{p}=0,
\end{equation}
where
\begin{equation}
H_p-\epsilon({\bf p})\mathds{1} =\left(\!\
  \begin{array}{cccc}
  -\ed \!&\!0\!&\! \tpd s_{x}\!&\! -\tpd s_y\\
  0\!&\!-\es - \tss z\!&\!\tsp s_{x}\!&\! \tsp s_{y} \\
  \tpd s_{x}\!&\! \tsp s_{x}\!&\!-\ep \!&\!-\tpp s_{x}s_{y} \\
  -\tpd s_{y}\!&\! \tsp s_{y}\!&\!-\tpp s_{x}s_{y}\!&\!-\ep
\end{array} \label{eq:hamil}
\!\right),
\end{equation}
and $\psi_p=(D_p,S_p,X_p,Y_p)$  is the wave function in momentum
space.

After some algebra the secular equation takes the form
\begin{equation}
\begin{array}{ll}
D({\bf p})=\!\!& \rm{det}(H_{p}-\epsilon({\bf
p})\mathds{1})\!=\!{\cal
A}xy + {\cal B}(x+y) + {\cal C} \\
&\\ &+\;z[{\cal K}xy + {\cal L}(x+y) +{\cal M}]=0 \label{eq:det}
\end{array}
\end{equation} with energy-dependent coefficients
\begin{equation}
\begin{array}{ll}
{\cal A}(\epsilon) = 16(4\tpd^2\tsp^2+2\tsp^2\tpp\ed -2\tpd^2\tpp\es-\tpp^2\ed\es ), \\
&\\{\cal B}(\epsilon) = -4\ep(\tsp^2\ed +\tpd^2\es),\;\; \;\;\; {\cal C}(\epsilon) = \ed\ep^2\es, \\
&\\{\cal K}(\epsilon) =  -\tss\tpp (\ed \tpp+2\tpd ^2),\;\;
 {\cal L}(\epsilon) = -\ep\tss\tpd^2,\;\; \\
&\\{\cal M}(\epsilon) = \ed\ep^2\tss,  \label{eq:ABC} \\
\end{array}
\end{equation}
where $\es = \epsilon({\bf p})-\eps,$ $\ep = \epsilon({\bf
p})-\epp,$ $\ed = \epsilon({\bf p})-\epd,$ are the energies taken
into account from single site atomic levels.
Due to the small numerical value of \tss\ the modulation of FS in
$p_z$ direction is also small.

\section{Analysis of the influence of the interlayer
hopping amplitude ${\rm {\bf{\tss}}}$}
%

Comparing our secular determinant with the purely 2D case
$\tss=0$, in (\ref{eq:det}) one can see that the influence of
interlayer hopping is formally reduced to a momentum dependent
single site energy shift for the Cu~4s level
\begin{equation}
\eps\rightarrow \eps - \tss z({\bf p}). \label{eq:ss}
\end{equation}
This is a diagonal matrix element whose influence is just zero on the
$p_y=\pi$ and $p_x=\pi$ lines in  2D Brillouin zone, shown in
figure~1, which we will discuss later. On the diagonal, where
$p_x=p_y$ and $p_y=2\pi-p_x$, the influence of this term is also
negligible. The 2D Hamiltonian (for $\tss=0$) has such
eigenfunctions~\cite{Mishonov}:
\begin{equation}
 \psi_p\!=\!\!\left(\!\!\
  \begin{array}{c}
    D_p \\
    S_p \\
    X_p \\
    Y_p \\
\end{array}
\!\!\right)=\!\left(\!\
  \begin{array}{c}
    -\es\ep^2 + 4 \ep\tsp^2(x+y) - 32 \tpp\tau^2_{sp}xy\\
        -4\ep\tsp\tpd(x-y) \\
        -(\es\ep - 8\tau^2_{\mathrm{sp}}y) \tpd s_x \\
         (\es\ep - 8\tau^2_{\mathrm{sp}}x) \tpd s_y
\end{array}\label{eq:psi}
\!\right),
 \end{equation}
where $\tau_{sp}^2=\tsp^2 - \frac{1}{2}\varepsilon_{s}\tpp$. After
the calculation of $\psi_p$ for $\epsilon=\epsilon^{(2D)}({\bf
p})$ we have to make the normalization $\psi_p:=\psi_p/\sqrt{D_
p^2+ S_p^2+X_p^2+Y_p^2}$. In first perturbational approximation
the influence of the diagonal term (\ref{eq:ss}) gives an addition
to the band energy $\epsilon({\bf p})$
\begin{equation}
\begin{array}{ll}
 W({\bf p}) = -8\tss \cos\left(\frac{p_x}{2}\right)
 \cos\left(\frac{p_y}{2}\right) \cos(p_z) S_p^2,\quad\\
 &\\
 \varepsilon({\bf p})= \varepsilon^{(2D)}({\bf p})+W({\bf p}),
 \label{eq:eps}
\end{array}
\end{equation}
i.e we present the band energy as a sum of a 2D band energy and a
correction by taking into account the hybridization in c-direction
W({\bf p}). Since $S_p\propto (x-y)$ this correction reduces to
zero on the diagonals of Brillouin zone and in such a way the
influence of \tss\ vanishes at 8 points of 2D Fermi contour at
every 45 degrees at horizontal, vertical and diagonal lines
crossing $(\pi,\pi)$ point. For the overdoped Tl:2201 the hole
pocket centered at $(\pi,\pi)$ point takes $62\%$ of the Brillouin
zone, see figure1.

In initial approximation $\tss=0$ we have a 2D secular equation
\begin{equation}
\rm{det}(H_{p}-\epsilon^{(2D)}({\bf p})\mathds{1})= {\cal A}xy +
{\cal B}(x+y) + {\cal C},
\end{equation}
which gives explicit expressions for the lower and upper arch of constant
energy contour (CEC) \cite{MishonovJoseph}
\begin{equation}
\begin{array}{ll}
p_{y}^{\rm (low)}(p_x) = 2\arcsin\sqrt{-\frac{{\cal B}x+{\cal C}}
{{\cal A}x+{\cal B}}},\quad\\
&\\ p_{y}^{\rm (up)}(p_x)=2\pi-p_{y}(p_x). \label{eq:CEC}
\end{array}
\end{equation}
For the band velocity at energy equal to Fermi energy
$\varepsilon^{(2D)}({\bf p})= E_F$
\begin{equation}
{\bf v }= \frac{\partial \varepsilon^{(2D)}({\bf p})} {\partial
{\bf p}},\qquad v_{F}=\sqrt{v_{x}^2 + v_{y}^2},
\end{equation}
we also have simple explicit expressions \cite{MishonovJoseph}
\begin{equation}
\begin{array}{ll}
v_{x}= -\frac{1}{2}\frac{(Ay+B)\sin(p_x)}{{\cal A}'xy+{\cal
B}'(x+y)+{\cal C}'},\qquad \\
&\\ v_{y}= -\frac{1}{2}\frac{(Ax+B)\sin(p_y)}{{\cal A}'xy+{\cal
B}'(x+y)+{\cal C}'}\label{eq:vel},
\end{array}
\end{equation}
\begin{equation}
\begin{array}{ll}
\label{velocity}
 v_{F} = \left|\frac{\partial \varepsilon^{(2D)}({\bf p})}{\partial \bf
 p}\right|\\
&\\ = \frac{
     \left[({\cal A}y+{\cal B})^2x(1-x)
               +({\cal A}x+{\cal B})^2y(1-y)\right]^{1/2}
   }{
     |{\cal A}'xy+{\cal B}'(x+y)+{\cal C}'|
   };
\end{array}
\end{equation}
the velocity in m/s is actually $a_0{\bf v}/\hbar.$ The
coefficients in the denominator ${\cal A}'$, ${\cal B}'$ and
${\cal C}'$ are energy derivatives of the polynomials
(\ref{eq:ABC}),
\begin{equation}
  \begin{array}{ll}
 {\cal A}'(\epsilon)&= 16\left[2\tsp^2\tpp-2\tpd^2\tpp-\tpp^2(\ed+\es)\right]={\rm d}{\cal A}/{\rm d}\epsilon,\\
 &\\ {\cal B}'(\epsilon)&= -4(\tsp^2\ed+\tpd^2\es) -4\ep(\tsp^2+\tpd^2)={\rm d}{\cal B}/{\rm d}\epsilon,\\
 &\\ {\cal C}'(\epsilon)&= \es\ep^2+\ed\ep^2+2\ed\es\ep={\rm d}{\cal C}/{\rm d}\epsilon.
\end{array}
\end{equation}
Our problem is to take into account the influence of the
perturbation (\ref{eq:eps}) to the Fermi contour (\ref{eq:CEC}).
Under the influence of the perturbation $W({\bf p})$ every point
${\bf p}$ of this CEC is shifted in a perpendicular to CEC direction
with momentum shift
\begin{equation}
\begin{array}{ll}
{\Delta {\bf p}}= -W({\bf p})\frac{\bf v}{v_{F}^2},\quad v_{F}=
\frac{|W({\bf p})|}{|\Delta {\bf p}|}, \\
&\\ \quad \varepsilon({\bf p})=
 \varepsilon^{(2D)}({\bf p})+{\bf v }\cdot{\Delta {\bf p}}.
\label{eq:v}
\end{array}
\end{equation}
The coordinates of the perturbed CEC are $p_x+{\Delta p_x}$ and
$p_y+{\Delta p_y};$ in such a way we approximatively built the 3D
Fermi surface. The formulae above represent a self-explainable
derivation: (1) the band energy is approximately presented by
gradient expansion (2) the ratio of energy difference and momentum
difference is equal to Fermi velocity (3) the shift of Fermi
contour in momentum space is parallel to the Fermi velocity.
Projections of this Fermi surfaces in 2D Brillouin zone are
depicted in figure 1. The LCAO approximation gives a similar
shape of the Fermi contours as the experimental Fermi surface
\cite{Hussey}. We have taken the set of parameters from
\cite{Andersen}: $\eps=6.5$~eV, $\epd=0$~eV, $\epp=-0.9$~eV,
$\tpd=1.6$~eV, $\tpp=0$~eV, $\tsp=2.3$~eV. The Fermi level
$E_F=1.89$~eV is determined to give $f=62\%$ hole filling

\cite{Hussey} of the 2D Brillouin zone
\begin{equation}
\frac{8}{(2\pi)^2} \int_{p_d}^\pi (p_x-p_y(p_x;E_F))\,{\rm
d}p_x=f,
\end{equation}
where $p_d$ is the solution of the equation
$0<p_y(p_d;E_F)=p_d<\pi$ \cite{MishonovPenev}
\begin{equation}
x_d=\sin^2\left(\frac{p_d}{2}\right)
=\frac{1}{\cal{A}}\left(-\cal{B} +\sqrt{ \cal{B}^{\rm
2}-\cal{A}\cal{C} } \right).
\end{equation}
The Fermi contour pases trough the points $D=(p_d,\,p_d)$ and
$C=(p_c,\,\pi),$ where
\begin{equation}
x_c=\sin^2\left(\frac{p_c}{2}\right)
=-\frac{\cal{B}+\cal{C}}{\cal{A}+\cal{B}},
\end{equation}
or $p_c=p_y(\pi;E_F).$ Finally interlayer hopping amplitude
$\tss=140$~meV is determined by the comparison with the experiment
of the modulation of FS~\cite{Hussey}.

\begin{figure}[t]
\centering
\includegraphics[width=0.6\columnwidth]{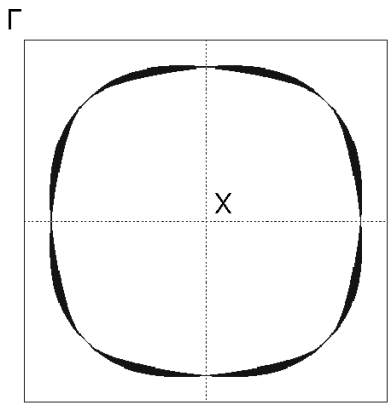}
\caption{Projections of 3D FS in 2D Brillouin zone obtained by
equation (\ref{eq:v}). In agreement with AMRO data \cite{Hussey}
the small modulation due to Cu~4s-Cu~4s tunnelling amplitude \tss\
vanishes at 8 highly symmetric lines passing through $(\pi,\pi)$
point. The hole pocket centered at $(\pi,\pi)$ point has
approximately $62\%$ of the surface of the 2D Brillouin zone. The warping is not to scale, it is exaggerated after \cite{Hussey} in order to emphasise its angular dependence which can be explained in the framework of the LCAO method.
\label{fig:1}}
\end{figure}

\newpage

\section{Comparison with ARPES data}
%

The galvanomagnetic phenomena such as AMRO are sensitive mainly to the
total area of the sections of the Fermi surface. The angle
resolved photoemission spectra (ARPES) however are sensitive to
the shape of the quasi 2D Fermi surface~\cite{Damascelli}. In
order to make a compromise conserving the area of the Fermi
surface, actually on the cross-section at $p_z=0,$ we can try to fit
its shape. The Fermi contour extracted from ARPES data for
Tl:2201~\cite{Plate:05} we can use the diagonal point
$\mathcal{D}=(0.3576\, \pi/a_0, 0.3576\, \pi/a_0)$ and another
point $\mathcal{C}=(0.1256\, \pi/a_0, \pi/a_0)$ as reference
points. We can start from the set of LCAO parameters given in
reference~\cite{Andersen} and changing only the Fermi level $E_F$ and
Cu~4s level $\epsilon_\mathrm{s}$ we can pass the Fermi contour
through the reference points $\mathcal{C}$ and $\mathcal{D}$ as it
is done in figure~2. The Fermi contour reproduces the shape from
reference~\cite{Plate:05}, its area is in agreement with the AMRO
data~\cite{Hussey}, and even the $E_F$ and $\epsilon_\mathrm{s}$
are not very different from the fit of LDA calculations by
Andersen \textit{et al.}~\cite{Andersen}.
\begin{figure}[t]
\centering
\includegraphics[width=0.6\columnwidth]{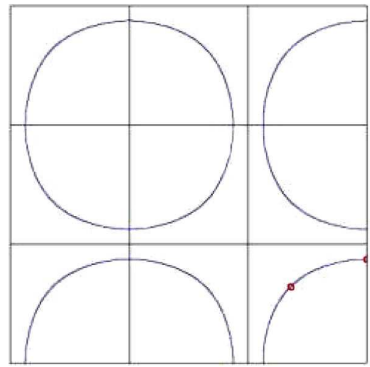}
\caption{A two dimensional section of Fermi surface for $p_z=0.$ The
theoretical Fermi contour is passing through the reference points in $k$-space: $\mathcal{C}=(0.1256\,
\pi/a_0, \pi/a_0)$ and $\mathcal{D}=(0.3576\, \pi/a_0, 0.3576\,
\pi/a_0)$ marked with small
(red) circles.
The short segment close to the saddle point $(\pi/a_0,0/a_0)$ is
the cut-III from Ref.~\cite{Plate:05} analyzed further at figure
3. \label{fig:2}}
\end{figure}

The fit of the absolute value of the energy scale however requires
a big compromise. As it is well-known the LDA often gives
overbinding of order 2 or even 3. Correcting overbinding in
local-density-approximation calculations we can insert an energy
renormalization scale so that the shape of the Fermi surface to be
exactly conserved and only the energy width along some well
investigated cut to coincide with the experiment. We used the
cut-III from reference~\cite{Plate:05} to fix the energy scale.
This cut is given as a short segment in figure 2. In figure 3 this
cut, energy versus quasimomentum, is given as leftmost segment.
The circles trough which the dispersion line passes are reference
points chosen from the experimental data~\cite{Plate:05}. These
reference points determine the energy scale.

On the right of this reference segment is presented the standard
energy dispersion along highly symmetric directions
(0,0)--($\pi,$0)--($\pi,$$\pi$)--(0,0) in momentum space. One of
the reference points was shifted in vertical direction in order
to use only the energy width of the cut-III for energy
determination.
\begin{figure}[t]
\centering
\includegraphics[width=0.6\columnwidth]{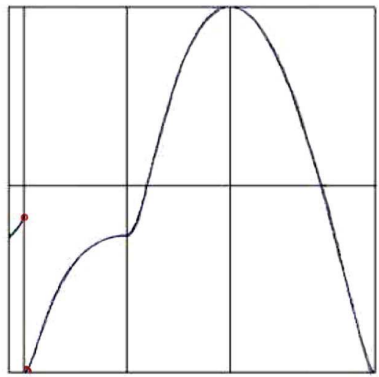}
\caption{Energy (in arbitrary units) as function of momentum along
the cut-III~\cite{Plate:05} presented in quasimomentum space in
figure~2. The two (red in the color version) circles are reference
points determining the vertical shift of the dispersion curve and
the renormalization of all LDA energy parameters. For comparison
are presented standard theoretical cuts $\varepsilon(\mathbf{p})$
for the triangle: (0,0)--($\pi,$0)--($\pi,$$\pi$)--(0,0).
\label{fig:3}}
\end{figure}
We believe that the so determined band structure has predictive
capabilities in a sense that we can use analytical formulae for the
band structure together with the fitted values of the LCAO
parameters in order to predict the results of new ARPES measurements.
In short, the mission of the theoretical physics is to predict the
results of experiments not done by nobody before. As a further
perspective we believe that single site energies have to be taken
from the experimental data processing of spectroscopic data for
interband transitions. Of course such an interpretation suppose
active use of the generic 4-band model taking into account the
accessories for every cuprate. Every additional experiment for
absolute determination of the energy scale can be an indispensable
tool for the final determination of the energy scale of LCAO
parameters. In the next section we propose a simple electronic
measurement which should be done with a layered
metal-insulator-superconductor structure with the same
superconductor.

\section{Determination of logarithmic derivative of
density of states by electronic measurements}

The analysis of Fermi surface by AMRO and ARPES as a rule leads only
to the determination of relative values of the parameters of the LCAO
Hamiltonian. In order to extract the absolute values we need to
fit band widths, cyclotron frequencies, density of states
$\nu(E_F)$ at Fermi energy $E_F$ or other quantities having
dimension of energy. We consider as very important the comparison of
different methods for investigations of band structure. The AMRO
gives the exact value of the area of sections of Fermi surface,
the ARPES is more sensitive for the shape and will be nice if this
knowledge is completed by some other method giving the exact value
of some energy parameter.
\begin{figure}[t]
\centering
\includegraphics[width=.85 \columnwidth]
{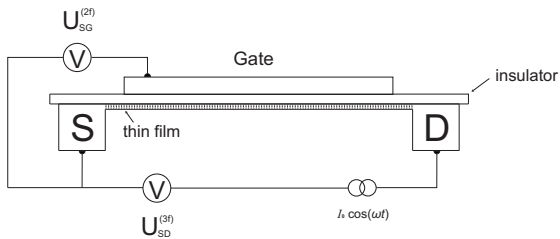} \caption{A field effect transistor(FET) is
schematically illustrated. The current $I(t)$, applied between the
source~(S) and the drain~(D) has a frequency $\omega$. Running
through the transistor the electrons create voltage
$U_{\mathrm{SG}}$ with double frequency $2\omega$ between the
source~(S) and the gate~(G). The source-drain voltage
$U_{\mathrm{SD}}$ is measured on the triple frequency $3\omega$.}
\label{fig:FET}
\end{figure}

The purpose of this section is to suggest a simple electronic
experiment, determining the logarithmic derivative of the density of
states (DOS) by electronic measurements using a thin film of the
investigated material Tl:2201. The proposed experiment requires the
preparation of field-effect transistor (FET) type microstructure and
requires standard electronic measurements. The FET controls the
current between two points but does so differently than the bipolar
transistor.  The FET relies on an electric field to control the
shape and hence the conductivity of a "channel" in a semiconductor
material. The shape of the conducting channel in a FET is altered
when a potential difference is applied to the gate terminal
(potential relative to either source or drain). It causes the
electrons flow to change it's width and thus controls the voltage
between the source and the drain. If the negative voltage applied to
the gate is high enough, it can remove all the electrons from the
gate and thus close the conductive channel in which the electrons
flow. Thus the FET is blocked.

The system, considered in this work is in hydrodynamic regime, which
means low frequency regime where the temperature of the
superconducting film adiabatically follows the dissipated Ohmic
power. All working frequencies of the lock-ins, say up to 100 kHz, are
actually low enough. The investigations of superconducting
bolometers show that only in the MHz range it is necessary to take into
account the heat capacity of the superconducting film. As an example
there is a publication corresponding to this topic
\cite{Mishonov:02}, as well as the references therein. In this work
we propose an experiment with a FET, for which we need to measure
the second harmonic of the source-gate voltage and the third
harmonic of the source-drain voltage. Other higher harmonics will be
present in the measurements (e.g. from the leads), but in principle
they can be also used for determination of the density of states. An
analogous experimental research has been already performed for
investigation of thermal interface resistance.\cite{Chenne:03} The
suggested experiment can be done using practically the same
experimental setup, only the gate electrodes should be added to the
protected by insulator layer superconducting films.

The purpose of this section is to suggest a simple electronic
experiment, determining the logarithmic derivative of the density of
states  by electronic measurements using a thin film of the material
Tl:2201. The thickness of the samples should be typical for the
investigation of high-Tc films, say 50-200 nm. Such films
demonstrate already the properties of the bulk phase. The numerical
value of this parameter
\be \nu \, '(E_F)=\frac{\mathrm{d} \nu(\epsilon)}{\mathrm{d}
\epsilon}, \ee
will ensure the absolute determination of the hopping integrals.

We propose a field effect transistor (FET) from Tl:2201
Fig.~{\ref{fig:FET}} to be investigated electronically with a lock-in
at second and third harmonics. Imagine a strip of Tl:2201 and
between the ends of the strip, between the source (S) and the drain
(D) an AC current is applied
\be \label{Eq:SD_current} I_\mathrm{SD}(t)=I_0\cos(\omega t). \ee

For low enough frequencies the ohmic power $P$ increases the
temperature of the film $T$ above the ambient temperature $T_0$
\be \label{Eq:power} P=RI_\mathrm{SD}^2=\alpha(T-T_0), \ee
where the constant $\alpha$ determines the boundary
thermo-resistance between the Tl:2201 film and the substrate, and
$R(T)$ is the temperature dependent source-drain (SD) resistance. We
suppose that for a thin film the temperature is almost homogeneous
across the thickness of the film. In such a way we obtain for the
temperature oscillations
\be \label{Eq:temperature}
(T-T_0)=\frac{RI_{SD}^2}{\alpha}=\frac{RI_0^2}{\alpha}\cos^2{(\omega
t)}. \ee
As the resistance is weakly temperature dependent
\be \label{Eq:resistance} R(T)=R_0 + (T-T_0) R_0\, ', \quad R_0\,
'(T_0)
    = \left.\frac{\mathrm{d} R(T)}{\mathrm{d} T}\right|_{T_0}.
\ee
A substitution here of the temperature oscillations from
Eq.~(\ref{Eq:temperature}) gives a small time variations of the
resistance
\be \label{Eq:SD_resistance} R(t) =
R_0\left(1+\frac{R_0'}{\alpha}I_0^2\cos^2(\omega t)\right). \ee
Now we can calculate the source-drain voltage as
\be U_{\mathrm{SD}}(t)=R(t)I_\mathrm{SD}(t). \ee
Substituting here the SD current from Eq.~(\ref{Eq:SD_current}) and
the SD resistance from Eq.~(\ref{Eq:SD_resistance}) gives us for the SD
voltage
\be \label{Eq:new_U_SD}
U_\mathrm{SD}(t)=U_\mathrm{SD}^{(1f)}\cos(\omega t)
  + U_\mathrm{SD}^{(3f)}\cos(3\omega t).
\ee
The coefficient in front of the first harmonic
$U_\mathrm{SD}^{(1f)}\approx R_0 I_0$ is determined by the SD
resistance $R_0$ at low currents $I_0,$ while for the third harmonic
signal using the elementary formula $\cos^3{(\omega
t)}=(3\cos{(\omega t)}+\cos{(3\omega t)})/4$ we obtain
\be \label{Eq:3f} U_{\mathrm{SD}}^{(3f)}=
\frac{U_{\mathrm{SD}}^{(1f)}}{4\alpha}I_0^2R_0'. \ee
From this formula we can express the boundary thermo-resistance by
electronic measurements
\be \label{Eq:alpha}
\alpha=\frac{U_{\mathrm{SD}}^{(1f)}}{4U_{\mathrm{SD}}^{(3f)}}I_0^2R_0'.
\ee

The realization of the method requires fitting of $R(T)$ and
numerical differentiation at working temperature $T_0;$ the linear
regression is probably the simplest method if we need to know only
one point.

At known $\alpha$ we can express the time oscillations of the
temperature substituting in Eq.~(\ref{Eq:temperature})
\begin{equation}
 \label{Eq:temp}
T\!=\!T_0\!+\!\frac{RI_0^2}{2\alpha}\left(1+\cos(2\omega t)\right)
\!\approx \! T_0\!\left(1\!+\!
  \frac{R_{\mathrm{SD}}I_0^2}{2\alpha T_0}
  \cos (2\omega t)\! \right)\!.
\end{equation}
In this approximation terms containing $I_0^4$ are neglected and
also we consider that the shift of the average temperature of the film
is small.

The variations of the temperature lead to a variation of the work
function of the film according to the well-known formula from the
physics of metals
\be \label{Eq:work-function} W(T)=-\frac{\pi^2}{6e}\frac{\nu \,
'}{\nu}k_B^2T^2, \quad \nu \, '(E_F)
 =\left.\frac{\mathrm{d} \nu}{\mathrm{d} \epsilon}\right|_{E_F},
\ee
where the logarithmic derivative of the density of states
$\nu(\epsilon)$ taken for the Fermi energy $E_F$ has dimension of
inverse energy, the work function $W$ has dimension of voltage, $T$
is the temperature in Kelvins and $k_B$ is the Boltzmann constant.
For an introduction see the standard textbooks on statistical
physics and physics of metals.\cite{Landau, Lifshitz} Substituting
here the temperature variations from Eq.~(\ref{Eq:temp}) gives
\be \label{Eq:wtt} W=-\frac{\pi^2 k_B^2}{6e} \frac{\nu \, '}{\nu}
 T_0^2 \left[1+\frac{R_0I_0^2}{\alpha T_0}\cos(2\omega t)\right]
 +\mathcal{O}(I_0^4),
\ee
where the $\mathcal{O}$-function again marks that the terms having
$I_0^4$ are negligible.

The oscillations of the temperature creates AC oscillations of the
source-gate (SG) voltage. We suppose that a lock-in with a
preamplifier, having high enough internal resistance is switched
between the source and the gate. In these conditions the second
harmonics of the work-function and of the SG voltage are equal
\be
\begin{array}{ll}
 U_\mathrm{SG}^{(2f)}= -\frac{\pi^2 k_B^2}{6e}
 \frac{\nu \,'}{\nu}T_0^2
 \frac{R_0I_0^2}{\alpha T_0},\\
  &\\ U_\mathrm{SG}(t)=U_\mathrm{SG}^{(2f)}\cos(2\omega t)
 +U_\mathrm{SG}^{(4f)}\cos(4\omega t) + \dots
\end{array}
\ee
Substituting $\alpha$ from Eq.~(\ref{Eq:alpha}) we have
\be \label{usgf} U_{\mathrm{SG}}^{(2f)}=-\frac{4\pi^2 k_B^2}{6e}
\frac{\nu{\,'}}{\nu}
\frac{U_{\mathrm{SD}}^{(3f)}}{I_0}\frac{T_0}{R_0'}. \ee
From this equation we can finally express the pursued logarithmic
derivative of the density of states
\be \label{density} \left. \frac{\mathrm{d}\ln
\nu(\epsilon)}{\mathrm{d}\epsilon}\right|_{E_F}=
\frac{\nu\,'}{\nu}=-\frac{3e}{2\pi^2k_B^2} \frac{I_0}{T_0}
\frac{U_{\mathrm{SG}}^{(2f)}}{U_{\mathrm{SD}}^{(3f)}}
\frac{d{R}}{d{T}}. \ee

In such a way the logarithmic derivative of the density of states can
be determined by fully electronic measurements with a FET. This
important energy parameter can be used for absolute determination of
the hopping integrals in the generic LCAO model. The realization of
the experiment can be considered as continuation of already
published detailed theoretical and experimental investigations and
having a set of complementary researches we can reliably determine
the LCAO parameters.

\section{Discussion and conclusions}
%
The suggested LCAO model for FS of Tl:2201 describes the important
qualitative properties of $p_z$ modulation: vanishing of this
modulation at 8 symmetric points of the 2D Fermi contours. This is an
important hint that \tss\ amplitude dominates in the formation of
coherent 3D Fermi surface and other interlayer tunnelling
amplitudes are irrelevant. This qualitative conclusion for the
importance of Cu~4s c-axis tunnelling matrix elements is in
agreement with the long-predicted analysis by Andersen et
al.~\cite{Andersen}. We wish to add a few words concerning the
pairing mechanism in Tl:2201 and cuprates in general. Wherever it
is hidden its influence has to be reduced to effective momentum
dependent scattering amplitude for electron pairs from conduction
3d band with opposite momentums; band structure created by
3d--2p--4s hybridization for which the revealing of FS is an
indispensable first step.\\

\noindent \textbf{Acknowledgements} One of the authors (TM) is
thankful to A.~Damascelli, N.~Hussey, and E.~Penev for the interest
to the paper, comments, technical help, suggestions and extra
details from their research. The authors are thankful to S.~Savova
for collaboration in the initial stages of this research related to
the preparation of the figures.

\newpage
\section{Appendix A: Calculation of eigenvector}

In the equation for eigenvectors $(H_p^{(2D)}-\epsilon({\bf
p})\mathds{1})\psi_{p}$, with $H_p^{(2D)}$ from (\ref{eq:hamil})
we search for a solution in the form $\psi_p=(1,S_p,X_p,Y_p)$. In
such a way we obtain the system
\begin{equation}
\left(\
  \begin{array}{ccc}
  -\es \!&\!\tsp s_x\!&\! \tsp s_y\\
  \tsp s_x\!&\!-\ep\!&\! -\tpp  s_x s_y \\
  \tsp s_y\!&\!-\tpp s_xs_y\!&\!-\ep
\end{array}
\right)\left(\
  \begin{array}{c}
    S \\
    X \\
    Y \\
\end{array}
\right)=\left(\
  \begin{array}{c}
    0 \\
    -\tpd s_x \\
    \tpd s_y \\
\end{array}
\right).
\end{equation}
The solution $S=\Delta_S/\Delta$, $X=\Delta_X/\Delta$,
$Y=\Delta_Y/\Delta$ is presented by the determinants

\begin{equation}
\Delta={\rm det}\left(\!\
  \begin{array}{ccc}
  -\es \!&\!\tsp s_x\!&\! \tsp s_y\\
  \tsp s_x\!&\!-\ep\!&\! -\tpp  s_x s_y \\
  \tsp s_y\!&\!-\tpp s_xs_y\!&\!-\ep
\end{array}
\!\right),
\end{equation}
\begin{equation}
 \Delta_S={\rm det}\left(\!\
  \begin{array}{ccc}
  0 \!&\!\tsp s_x\!&\! \tsp s_y\\
  -\tpd s_x\!&\!-\ep\!&\! -\tpp  s_x s_y \\
  \tpd s_y\!&\!-\tpp s_xs_y\!&\!-\ep
\end{array}
\!\right),
\end{equation}
\begin{equation}
\Delta_X={\rm det}\left(\!\
  \begin{array}{ccc}
  -\es \!&\!0\!&\! \tsp s_y\\
  \tsp s_x\!&\!-\tpd s_x\!&\! -\tpp  s_x s_y \\
  \tsp s_y\!&\!\tpd s_y\!&\!-\ep
\end{array}
\!\right),
\end{equation}
\begin{equation}
\Delta_Y={\rm det}\left(\!\
  \begin{array}{ccc}
  -\es \!&\!\tsp s_x\!&\! 0\\
  \tsp s_x\!&\!-\ep\!&\! -\tpd s_x\\
  \tsp s_y\!&\!-\tpp s_xs_y\!&\!\tpd s_y
\end{array}
\!\right).
\end{equation}
 Multiplication by $\Delta$ gives the eigen vector
$(D,S,X,Y)=(\Delta,\Delta_S,\Delta_X,\Delta_Y)$ presented in
equation (\ref{eq:psi}).

\end{document}